\documentclass{article}
\usepackage{spconf, amsmath, graphicx, amsfonts, amssymb, amsthm, float}
\title{Linearly Reconfigurable Kalman Filtering for a Vector Process}
\name{Feng Jiang, Jie Chen, and A. Lee Swindlehurst}
\address{Department of Electrical Engineering and Computer Science\\
  University of California at Irvine,\\ Irvine, CA 92697, USA\\
  Email:\{feng.jiang, jie.chen, swindle\}@uci.edu}

\begin{document}
%\ninept
%
\maketitle
\begin{abstract}
In this paper, we consider a dynamic linear system in state-space form where the observation equation depends linearly on a set of parameters.  We address the problem of how to dynamically calculate these parameters in order to minimize the mean-squared error (MSE) of the state estimate achieved by a Kalman filter.  We formulate and solve two kinds of problems under a quadratic constraint on the observation parameters: minimizing the sum MSE (Min-Sum-MSE) or minimizing the maximum MSE (Min-Max-MSE). In each
case, the optimization problem is divided into two sub-problems for which optimal solutions can be found: a semidefinite programming (SDP) problem followed by a constrained least-squares minimization.  A more direct solution is shown to exist for the special case of a scalar observation; in particular, the Min-Sum-MSE problem is optimally solved utilizing Rayleigh quotient, and the Min-Max-MSE problem reduces to an SDP feasibility test that can be solved via the bisection method.
\end{abstract}

\setlength{\baselineskip}{0.95\baselineskip}
\section{Introduction}
Dynamic state-space models in which the observation equation depends on parameters that can be adaptively tuned to improve performance have recently been proposed by several authors.  For example, in \cite{Salmi:2009}, dynamic wireless channel parameters such as the delay of arrival, the angle of arrival, the angle of departure, etc, are tracked via a Kalman filter whose performance depends on properties of the antenna array.  In \cite{Zhan:2010}, the parameters to be estimated are the position and velocity of a target and the observations at a set of mobile sensors are the time delay and Doppler shift of the signal reflected by the target.  The positions of the mobile sensors are adjusted in order to minimize the tracking error of a standard extended Kalman filter.  In \cite{Jiang:2012}, a distributed sensor network problem is considered where the observed signal is a linear function of the transmission gain of each sensor, and optimal values for these gains are found under different power constraints to minimize the mean-squared error (MSE) of a scalar variable at the fusion center.

In this paper, we consider a scenario that generalizes the one assumed in \cite{Jiang:2012} by allowing the estimated parameters and the observations to be vector- rather than scalar-valued.  The state-space observation matrix is assumed to depend linearly on a set of parameters, and we consider the problem of optimizing these parameters in order to minimize the MSE obtained by a Kalman filter that tracks the unknown state.  Two different optimization problems are considered: one that minimizes the sum MSE (Min-Sum-MSE) over all the parameters, and another that minimizes the maximum MSE (Min-Max-MSE) of all parameters.  In the general case, we divide the overall problem into two sub-problems whose optimal solutions can be found.  The first sub-problem estimates the optimal observation matrix without taking the linear structure into account, and the second finds the set of parameters that are closest to the resulting observation matrix under a quadratic constraint on the parameters themselves.  Dividing the problem into these two steps will cause a performance loss, but simulations demonstrate that the loss is minimal, and performance is close to the lower bound given by the solution to the unconstrained problem.  We also consider the special case of a scalar observation, and show that in this case the Min-Sum-MSE problem is converted to a Rayleigh quotient maximization problem, for which an optimal closed-form solution is obtained, and we show that for the Min-Max-MSE problem, a relaxed version of the problem leads to a simple SDP feasibility test that can be solved via the bisection algorithm.  Simulation results show that in most cases, the solution to the relaxed and unrelaxed problems are the same.

\iffalse
The rest of this paper is organized as follows. Section \ref{sec:two} describes the signal model for the vector observation case and formulates the sum MSE minimization problem and maximum MSE minimization problem.  The system model and problem formulation for the scalar observation case are presented in Section \ref{sec:three}. Numerical results are then provided in Section \ref{sec:four} and the conclusions are summarized in Section \ref{sec:five}.\fi
\iffalse
Each problem is divided into two sub-problems which are solved using SDP and least square method respectively.\fi

\section{Vector Observation Model}\label{sec:two}
We assume the dynamic parameter to be estimated is a complex-valued vector that obeys
the following state-space model:
\begin{eqnarray}
\boldsymbol{\theta}_{n+1}&=&\mathbf{F}\boldsymbol{\theta}_n+\mathbf{u}_n,\nonumber
%\mathbf{y}_k&=&f(\mathbf{a},\mathbf{\theta}_k)+\mathbf{v}_k\\
%y_k&=&f_1(\mathbf{a})\theta_k+f_2(\mathbf{a})v_k
\end{eqnarray}
where $\boldsymbol{\theta}_{n+1}\in\mathbb{C}^{M\times 1}$ is the parameter at time step $n+1$, $\mathbf{F}\in\mathbb{C}^{M\times M}$ is the state transition matrix, and $\mathbf{u}_n\sim\mathcal{CN}(\mathbf{0},\mathbf{Q})$ is the process noise. The observed signal vector is given by
\iffalse
\begin{eqnarray}
f_1(\mathbf{a})\approx f_1(\tilde{\mathbf{a}})+\frac{d f_1(\tilde{\mathbf{a}})}{d \mathbf{a}}(\mathbf{a}-\tilde{\mathbf{a}})\\
f_2(\mathbf{a})\approx f_2(\tilde{\mathbf{a}})+\frac{d f_2(\tilde{\mathbf{a}})}{d \mathbf{a}}(\mathbf{a}-\tilde{\mathbf{a}})
\end{eqnarray}

\begin{eqnarray}
y_k\approx (\mathbf{d}_1^T\mathbf{a}+c_1)\theta_k+(\mathbf{d}_2^T\mathbf{a}+c_2)v_k,
\end{eqnarray}

\begin{eqnarray}\label{eq:opt3}
\max_{\mathbf{a}}&&\frac{\mathbf{a}^T\mathbf{d}_1\mathbf{d}_1^T\mathbf{a}+2c_1\mathbf{d}_1^T\mathbf{a}+c_1^2}{\mathbf{a}^T\mathbf{d}_2\mathbf{d}_2^T\mathbf{a}+c_2^2}\\
s. t. &&\mathbf{a}^T\mathbf{D}\mathbf{a}=P_{\max}\; . \nonumber
\end{eqnarray}

update
\begin{eqnarray}
\tilde{\mathbf{a}}=\mathbf{a}^{*}
\end{eqnarray}
\fi
\begin{equation}
\mathbf{y}_n=\mathbf{C}\boldsymbol{\theta}_n+\mathbf{v}_n,\nonumber
\end{equation}
where $\mathbf{v}_n\sim\mathcal{CN}(\mathbf{0},\sigma_v^2\mathbf{I}_L)$ is the observation noise, $\mathbf{I}_L$ is the $L\times L$ identity matrix and $\mathbf{C}\in\mathbb{C}^{L\times M}$ is the observation matrix. We assume that $\mathbf{C}$ is a linear function of some parameters $\mathbf{a}\in\mathbb{C}^{N \times 1}$ such that $\mathrm{vec}[\mathbf{C}]=\mathbf{G}\mathbf{a}$ for a given $\mathbf{G}\in\mathbb{C}^{LM\times N}$.

The MSE of the state estimate is found via the standard Kalman filtering equations \cite{Kay:1993}:
\begin{itemize}
\item Prediction MSE Matrix
\begin{equation}
\mathbf{M}_{n|n-1}=\mathbf{F}\mathbf{M}_{n-1|n-1}\mathbf{F}^H+\mathbf{Q}\nonumber
\end{equation}
\item Kalman Gain Matrix
\begin{equation}
\mathbf{K}_n=\mathbf{M}_{n|n-1}\mathbf{C}^H(\sigma_v^2\mathbf{I}_L+\mathbf{C}\mathbf{M}_{n|n-1}\mathbf{C}^H)^{-1}\nonumber
\end{equation}
\item MSE matrix
\begin{eqnarray}
\mathbf{M}_{n|n}&=&(\mathbf{I}_M-\mathbf{K}_n\mathbf{C})\mathbf{M}_{n|n-1}\nonumber\\
&=&\bigg(\mathbf{M}_{n|n-1}^{-1}+\frac{1}{\sigma_v^2}\mathbf{C}^H\mathbf{C}\bigg)^{-1}\;.\nonumber
\end{eqnarray}
\end{itemize}

\subsection{Minimize Sum MSE}
In this section we consider the problem of minimizing the sum-MSE under a quadratic constraint of $\mathbf{a}$. The ideal optimization problem is formulated as
\begin{eqnarray}\label{eq:opt11}
\min_{\mathbf{a}} && \mathrm{tr}(\mathbf{M}_{n|n})\\
s. t. &&\|\mathbf{C(\mathbf{a})}\|_F^2\le P \; . \nonumber
\end{eqnarray}
The solution to~(\ref{eq:opt11}) is difficult to obtain directly, so instead we divide the optimization problem into two subproblems. We first find an unconstrained $\mathbf{C}^{*}$ that minimizes $\mathrm{tr}(\mathbf{M}_{n|n})$, and then based on $\mathbf{C}^{*}$, we obtain the approximate solution $\mathbf{a}^{*}$. 

The first step is to solve
\begin{eqnarray}\label{eq:opt1}
\min_{\mathbf{C}} && \mathrm{tr}(\mathbf{M}_{n|n})\\
s. t. &&\|\mathbf{C}\|_F^2\le P\nonumber\;.
\end{eqnarray}
Defining $\tilde{\mathbf{C}}=\mathbf{C}^H\mathbf{C}$, we can rewrite~(\ref{eq:opt1}) as
\begin{eqnarray}\label{eq:opt2}
\min_{\tilde{\mathbf{C}}} && \mathrm{tr}(\mathbf{D})\\
s. t. &&\mathbf{D}^{-1}=\mathbf{M}_{n|n-1}^{-1}+\frac{1}{\sigma_v^2}\tilde{\mathbf{C}}\nonumber\\
&&\mathrm{tr}(\tilde{\mathbf{C}})\le P\nonumber\\
&&\tilde{\mathbf{C}}\succeq 0\nonumber\;.
\end{eqnarray}
Replacing the equality in the first constraint of (\ref{eq:opt2}) with an inequality yields an equivalent optimization problem:
\begin{eqnarray}\label{eq:opt3}
\min_{\tilde{\mathbf{C}},\mathbf{D}} && \mathrm{tr}(\mathbf{D})\\
s. t. &&\mathbf{D}^{-1}\preceq\mathbf{M}_{n|n-1}^{-1}+\frac{1}{\sigma_v^2}\tilde{\mathbf{C}}\nonumber\\
&&\mathrm{tr}(\tilde{\mathbf{C}})\le P\nonumber\\
&&\tilde{\mathbf{C}}\succeq 0\nonumber\;.
\end{eqnarray}
The problem in (\ref{eq:opt3}) is equivalent to (\ref{eq:opt2}) in the sense that for the optimal solution of problem (\ref{eq:opt3}) the equality of the first constraint must hold. According to the Schur complement \cite{Luo:2004}, the first constraint in (\ref{eq:opt3}) is equivalent to :
\begin{eqnarray}\label{eq:schur}
\left[\begin{array}{cc}\mathbf{M}_{n|n-1}^{-1}+\frac{1}{\sigma_v^2}\tilde{\mathbf{C}}&\mathbf{I}_M\\
                \mathbf{I}_M&\mathbf{D}
                \end{array}\right]\succeq 0\;.
\end{eqnarray}
Plugging (\ref{eq:schur}) into (\ref{eq:opt3}), we have
\begin{eqnarray}\label{eq:opt4}
\min_{\tilde{\mathbf{C}},\mathbf{D}} && \mathrm{tr}(\mathbf{D})\\
s. t. &&\left[\begin{array}{cc}\mathbf{M}_{n|n-1}^{-1}+\frac{1}{\sigma_v^2}\tilde{\mathbf{C}}&\mathbf{I}_M\\
                \mathbf{I}_M&\mathbf{D}
                \end{array}\right]\succeq 0\nonumber\\
&&\mathrm{tr}(\tilde{\mathbf{C}})\le P\nonumber\\
&&\tilde{\mathbf{C}}\succeq 0\nonumber\;.
\end{eqnarray}
\iffalse
By converting the constraints of problem (\ref{eq:opt4}) into a large block diagonal linear matrix inequality, the above problem can be written in the standard form of an SDP:
\begin{eqnarray}\label{eq:standard}
\min_{\mathbf{\tilde{C}},\mathbf{D}} && \mathrm{tr}(\mathbf{D})\\
s. t. &&\left[\begin{array}{cccc}\mathbf{M}_{n|n-1}^{-1}+\frac{1}{\sigma_v^2}\tilde{\mathbf{C}}&\mathbf{I}&&\\
                \mathbf{I}&\mathbf{D}&&\\
                &&\tilde{\mathbf{C}}&\\
                &&&P-\mathrm{tr}(\tilde{\mathbf{C}})
                \end{array}\right]\succeq 0\nonumber\;.
\end{eqnarray}\fi
By converting the constraints of problem (\ref{eq:opt4}) into a large block diagonal linear matrix inequality, we can transform the problem into a standard SDP form, which can be efficiently solved using the interior point method. 

Denote the optimal solution to problem (\ref{eq:opt4}) as $\tilde{\mathbf{C}}^{*}$, and define the singular value decomposition of $\tilde{\mathbf{C}}^{*}$ as $\tilde{\mathbf{C}}^{*}=\mathbf{U}\mathbf{\Sigma}\mathbf{U}^H$, so that $\mathbf{C}^*=\mathbf{\Sigma}^{\frac{1}{2}}\mathbf{U}^{H}$. The performance of $\mathbf{C}^{*}$ provides a lower bound for problem (\ref{eq:opt11}). To estimate $\mathbf{a}^*$, we solve\footnote{At the optimal solution of (\ref{eq:opt11}), the constraint should attain equality. \iffalse When $\mathbf{G}$ is a fat matrix, there exists infinite number of solutions for $\mathbf{a}$ that lead to the optimal $C^{*}$.\fi}
\begin{eqnarray}\label{eq:lsform}
\min_{\mathbf{a}} && \|\mathrm{vec}(\mathbf{C}^*)-\mathbf{G}\mathbf{a}\|_2^{2}\nonumber\\
s. t. &&\mathbf{a}^H\mathbf{G}^{H}\mathbf{G}\mathbf{a}=P\nonumber\;,
\end{eqnarray}
which directly leads to 
\begin{equation}\label{eq:ls}
\mathbf{a}^{*}=\gamma(\mathbf{G}^H\mathbf{G})^{-1}\mathbf{G}^H\mathrm{vec}(\mathbf{C}^*)\;,
\end{equation}
where $\gamma$ is defined as $\gamma=\sqrt{\frac{P}{\mathrm{vec}(\mathbf{C}^*)^H\mathbf{G}(\mathbf{G}^H\mathbf{G})^{-1}\mathbf{G}^H\mathrm{vec}(\mathbf{C}^*)}}$.

\subsection{Minimize the Maximum MSE}
When the maximum MSE is to be minimized, the parameter optimization problem can be stated as
\begin{eqnarray}\label{eq:minmax}
\min_{\mathbf{a}}\max_{i} && [\mathbf{M}_{n|n}]_{i,i}\\
s. t. &&\|\mathbf{C(\mathbf{a})}\|_F^2\le P\;\nonumber. 
\end{eqnarray}
Similar to (\ref{eq:opt11}), when treating $\mathbf{C}$ as the variable to be optimized, we can rewrite problem (\ref{eq:minmax}) as
\begin{eqnarray}\label{eq:minmax2}
\min_{\mathbf{C}}\max_{i} && [\mathbf{M}_{n|n}]_{i,i}\\
s. t. &&\|\mathbf{C}\|_F^2\le P\nonumber\;.
\end{eqnarray}
Introducing an auxiliary variable $t$, we can rewrite (\ref{eq:minmax2}) as
\begin{eqnarray}\label{eq:minmax3}
\min_{\mathbf{C},t}   && t \\
s. t. &&t\ge [\mathbf{M}_{n|n}]_{i,i}\;,\nonumber\\
&&\|\mathbf{C}\|_F^2\le P\nonumber\;.
\end{eqnarray}
Define $\mathbf{e}_i$ as the vector with all zeros except for a 1 in the $i$th position, 
so that (\ref{eq:minmax3}) is equivalent to
\begin{eqnarray}\label{eq:minmax4}
\min_{\mathbf{C},t}   && t \\
s. t. &&t\ge \mathbf{e}_i^T\mathbf{M}_{n|n}\mathbf{e}_i\;,\nonumber\\
&&\|\mathbf{C}\|_F^2\le P\nonumber\;.
\end{eqnarray}
Again, we utilize the Schur complement to rewrite the first constraint in (\ref{eq:minmax4}) and we have
\begin{eqnarray}\label{eq:minmax5}
\min_{\tilde{\mathbf{C}},t}   && t \\
s. t. &&\left[\begin{array}{cc}t&\mathbf{e}_i\nonumber\\
                \mathbf{e}_i^T&\mathbf{M}_{n|n-1}^{-1}+\frac{1}{\sigma_v^2}\tilde{\mathbf{C}}
                \end{array}\right]\succeq 0,\;i=1,\cdots,N\nonumber\\
&&\mathrm{tr}(\tilde{\mathbf{C}})\le P\nonumber\\
&&\tilde{\mathbf{C}}\succeq 0\nonumber\;
\end{eqnarray}
Similar to (\ref{eq:opt4}), we can write the constraints of problem (\ref{eq:minmax5}) in a large block diagonal linear matrix inequality and convert the problem to a standard SDP form. After obtaining $\tilde{\mathbf{C}}^{*}$, we can use (\ref{eq:ls}) to find the solution $\mathbf{a}^{*}$.

\iffalse
\section{Application: Pilot Optimization for Channel Tracking of OFDM System}
\subsection{Channel Model}
\begin{equation}
h(t)=\sum_{l=1}^{L}h_l(t-lT_s-\tau_d)
\end{equation}
\begin{equation}
\mathbf{h}=[h_1,\dots,h_L]^T
\end{equation}
Dynamic Model
\begin{equation}
\mathbf{h}_n=\mathbf{A}\mathbf{h}_{n-1}+\mathbf{g}_n
\end{equation}
\begin{equation}
\mathbf{x}_n=\mathbf{B}\mathbf{F}\mathbf{h}_n+\mathbf{n}
\end{equation}

$\mathbf{B}$ is diagonal matrix with pilot symbols and optimize $\mathbf{B}$ to minimize estimation error.
\fi
%An potential application of the vector-observation model is the power allocation for the pilot %of OFDM system. In \cite{Larsen:2011}, the static channel model was investigated and when %assuming the multi-path channel dynamically changes according to the Gauss-Markov process, the %pilot can be estimated using the Kalman filter and the parameter $\mathbf{a}$ can be treated %as the power allocation for different pilots.

\section{Scalar Observation Model}\label{sec:three}
For a scalar observation, we have
\begin{equation}
y_n=\mathbf{c}^H\boldsymbol{\theta}_n+v_n,\nonumber
\end{equation}
where $\mathbf{c}=\mathbf{G}\mathbf{a}$ and as before $\mathbf{G}\in\mathbb{C}^{M\times N}$. The MSE of the estimated state is given by
\begin{equation}\label{eq:msesc}
\mathbf{M}_{n|n}=\mathbf{M}_{n|n-1}-\frac{\mathbf{M}_{n|n-1}\mathbf{c}\mathbf{c}^H\mathbf{M}_{n|n-1}}{\sigma_n^2+\mathbf{c}^H\mathbf{M}_{n|n-1}\mathbf{c}}.
\end{equation}

\subsection{Minimize Sum MSE}
For a scalar observation, the sum MSE is computed as
\begin{equation}
\mathrm{tr}(\mathbf{M}_{n|n})=\mathrm{tr}(\mathbf{M}_{n|n-1})-\frac{\mathbf{c}^H\mathbf{M}_{n|n-1}^2\mathbf{c}}{\sigma_n^2+\mathbf{c}^H\mathbf{M}_{n|n-1}\mathbf{c}}\;,\nonumber
\end{equation}
and we formulate the following optimization problem:
\begin{eqnarray}\label{eq:scalaropt1}
\max_{\mathbf{a}} && \frac{\mathbf{a}^H\mathbf{G}^H\mathbf{M}_{n|n-1}^2\mathbf{G}\mathbf{a}}{\sigma_n^2+\mathbf{a}^H\mathbf{G}^H\mathbf{M}_{n|n-1}\mathbf{G}\mathbf{a}}\\
s. t. &&\mathbf{a}^H\mathbf{G}^H\mathbf{G}\mathbf{a}\le P\nonumber\;.
\end{eqnarray}
Since the objective function in (\ref{eq:scalaropt1}) is monotonically increasing with the norm of $\mathbf{c}$, the constraint must be active at the optimal solution and we can rewrite the problem as
\begin{eqnarray}
\max_{\mathbf{a}} && \frac{\mathbf{a}^H\mathbf{G}^H\mathbf{M}_{n|n-1}^2\mathbf{G}\mathbf{a}}{\mathbf{a}^H(\frac{\sigma_n^2}{P}\mathbf{\mathbf{G}^H\mathbf{G}}+\mathbf{G}^H\mathbf{M}_{n|n-1}\mathbf{G})\mathbf{a}}\nonumber\\
s. t. &&\mathbf{a}^H\mathbf{G}^H\mathbf{G}\mathbf{a}=P\nonumber\;.
\end{eqnarray}
The solution to the above problem can be found directly as
\begin{equation}
\mathbf{c}^{*}=\sqrt{\frac{P}{\mathbf{u}^H\mathbf{B}\mathbf{u}}}\mathbf{B}^{-\frac{1}{2}}\mathbf{u},\nonumber
\end{equation}
where $\mathbf{B}=\frac{\sigma_n^2}{P}\mathbf{G}^H\mathbf{G}+\mathbf{G}^H\mathbf{M}_{n|n-1}\mathbf{G}$, and $\mathbf{u}$ is the eigenvector corresponding to the largest eigenvalue of  $\mathbf{B}^{-\frac{1}{2}}\mathbf{G}^H\mathbf{M}_{n|n}^2\mathbf{G}\mathbf{B}^{-\frac{1}{2}}$.

\subsection{Minimize Maximum MSE}

When the maximum MSE is to be minimized, the optimization problem becomes
\begin{eqnarray}\label{eq:minmaxsc}
\min_{\mathbf{a}}\max_{i} && [\mathbf{M}_{n|n}]_{i,i}\\
s. t. &&\mathbf{a}^H\mathbf{G}^H\mathbf{G}\mathbf{a}\le P\nonumber\;.
\end{eqnarray}
Substituting (\ref{eq:msesc}) into (\ref{eq:minmaxsc}), we have
\begin{eqnarray}\label{eq:minmaxsc2}
\min_{\mathbf{a}}\max_{i}&&\!\!\!\!\!\!\! [\mathbf{M}_{n|n-1}]_{i,i}-\frac{\mathbf{e}_i^T\mathbf{M}_{n|n-1}\mathbf{G}\mathbf{a}\mathbf{a}^H\mathbf{G}^H\mathbf{M}_{n|n-1}\mathbf{e}_i}{\sigma_n^2+\mathbf{a}^H\mathbf{G}^H\mathbf{M}_{n|n-1}\mathbf{G}\mathbf{a}}\nonumber\\
s. t. &&\mathbf{a}^H\mathbf{G}^H\mathbf{G}\mathbf{a}\le P\nonumber\;.
\end{eqnarray}
Introducing the auxiliary variable $t$, we have
\begin{eqnarray}\label{eq:minmaxsc3}
\min_{\mathbf{a}}\!\!\!\!\!&& t\\
s. t.\!\!\!\!\!&& [\mathbf{M}_{n|n-1}]_{i,i}-\frac{\mathbf{e}_i^T\mathbf{M}_{n|n-1}\mathbf{G}\mathbf{a}\mathbf{a}^H\mathbf{G}^H\mathbf{M}_{n|n-1}\mathbf{e}_i}{\sigma_n^2+\mathbf{a}^H\mathbf{G}^{H}\mathbf{M}_{n|n-1}\mathbf{G}\mathbf{a}}\le t, \nonumber\\
&&\hspace{13em}i=1,\cdots,N\nonumber\\
&&\mathbf{a}^H\mathbf{G}^H\mathbf{G}\mathbf{a}\le P\nonumber\;.
\end{eqnarray}
\iffalse
\begin{eqnarray}\label{eq:opt1}
\min_{\mathbf{a}}&& t\\
s. t.&& \mathbf{e}_i^T\mathbf{M}_{n|n}\mathbf{e}_i-\frac{\mathbf{e}_i^T\mathbf{M}_{n|n-1}\mathbf{c}\mathbf{c}^H\mathbf{M}_{n|n-1}\mathbf{e}_i}{\sigma_n^2+\mathbf{c}^H\mathbf{M}_{n|n-1}\mathbf{c}}\le t\\
&&\mathbf{c}^H\mathbf{c}\le P\nonumber
\end{eqnarray}

\begin{eqnarray}\label{eq:opt1}
\min_{\mathbf{a}}&& t\\
s. t.&& (\mathbf{e}_i^T\mathbf{M}\mathbf{e}_i-t)(\sigma_n^2+\mathbf{c}^H\mathbf{M}_{n|n-1}\mathbf{c})\le\mathbf{c}^H\mathbf{M}_{n|n-1}\mathbf{e}_i\mathbf{e}_i^T\mathbf{M}_{n|n-1}\mathbf{c}\\
&&\mathbf{c}^H\mathbf{c}\le P\nonumber
\end{eqnarray}\fi
After some mathematical manipulation, we can rewrite problem (\ref{eq:minmaxsc3}) into the following form
\begin{eqnarray}\label{eq:minmaxsc4}
\min_{\mathbf{a}}&& t\\
s. t.&& \big([\mathbf{M}_{n|n-1}]_{i,i}-t\big)\sigma_n^2\le\mathbf{a}^H\mathbf{E}_i\mathbf{a},\; i=1,\cdots,N\nonumber\\
&&\mathbf{a}^H\mathbf{G}^H\mathbf{G}\mathbf{a}\le P\nonumber,
\end{eqnarray}
where $\mathbf{E}_i$ is defined as $\mathbf{E}_i=\mathbf{G}^H(\mathbf{M}_{n|n-1}\mathbf{e}_i\mathbf{e}_i^T\mathbf{M}_{n|n-1}-([\mathbf{M}_{n|n-1}]_{i,i}-t)\mathbf{M}_{n|n-1})\mathbf{G}$.
Problem (\ref{eq:minmaxsc4}) is equivalent to
\begin{eqnarray}\label{eq:minmaxsc5}
\min_{\mathbf{a}}&& t\nonumber\\
s. t.&& \big([\mathbf{M}_{n|n-1}]_{i,i}-t\big)\sigma_n^2\le\mathrm{tr}\big(\mathbf{A}\mathbf{E}_i\big),\;i=1,\cdots,N\;\nonumber\\
&&\mathrm{tr}(\mathbf{A}\mathbf{G}^H\mathbf{G})\le P\;\nonumber\\
&&\mathrm{rank}(\mathbf{A})=1\;\nonumber\\
&&\mathbf{A}\succeq 0\;.\nonumber
\end{eqnarray}
At this point, we see that the problem could be efficiently solved without the rank constraint.  Consequently, we relax this constraint to yield a quasi-convex problem:
\begin{eqnarray}\label{eq:minmaxsc6}
\min_{\mathbf{a}}\!\!\!&& t\\
s. t.\!\!\!&& \big([\mathbf{M}_{n|n-1}]_{i,i}-t\big)\sigma_n^2\le\mathrm{tr}\big(\mathbf{A}\mathbf{E}_i\big),\; i=1,\cdots,N\;\nonumber\\
\!\!\!&&\mathrm{tr}(\mathbf{A}\mathbf{G}^H\mathbf{G})\le P\nonumber\\
\!\!\!&&\mathbf{A}\succeq 0\;.\nonumber
\end{eqnarray}
In the above problem, given a deterministic $t$, all the constraints are convex. Denote the optimal value of problem (\ref{eq:minmaxsc6}) as $t^{*}$, so that for a $\tilde{t}$ that makes the problem (\ref{eq:minmaxsc6}) feasible, we have $t^{*}\le \tilde{t}$, while if the problem is infeasible, we have $t^{*}>\tilde{t}$. To find $t^{*}$, we search over $t$ using the bisection method \cite{Havary-Nassab:2008}. For a given $t$, we solve the SDP feasibility problem:
\begin{eqnarray}\label{eq:minmaxsc7}
\mathrm{find} &&\mathbf{A}\nonumber\\
s. t.&& \big([\mathbf{M}_{n|n-1}]_{i,i}-t\big)\sigma_n^2\le\mathrm{tr}\big(\mathbf{A}\mathbf{E}_i\big),\; i=1,\cdots,N\;\nonumber\\
&&\mathrm{tr}(\mathbf{A}\mathbf{G}^H\mathbf{G})\le P\nonumber\;\\
&&\mathbf{A}\succeq 0 \nonumber
\end{eqnarray}
to obtain the optimal $t^{*}$, then we plug this $t^{*}$ into problem~(\ref{eq:minmaxsc6}) to find $\mathbf{A}^{*}$.  If $\mathrm{rank}(\mathbf{A}^{*})=1$, then $\mathbf{A}^{*}$ is the optimal solution to problem (\ref{eq:minmaxsc}), otherwise, a rank-one solution $\mathbf{a}^{*}$ can be reconstructed. The optimal value $t^{*}$ of problem~(\ref{eq:minmaxsc6}) can be used as a lower bound for the minimum MSE of problem (\ref{eq:minmaxsc}).  Our simulations indicate that in most cases, the rank of $\mathbf{A}^{*}$ is one, which indicates the performance of $\mathbf{a}^{*}$ is very close to the lower bound provided by $\mathbf{A}^{*}$.
\begin{figure}[!htb]
\centering
\includegraphics[height=2.5in, width=3.3in]{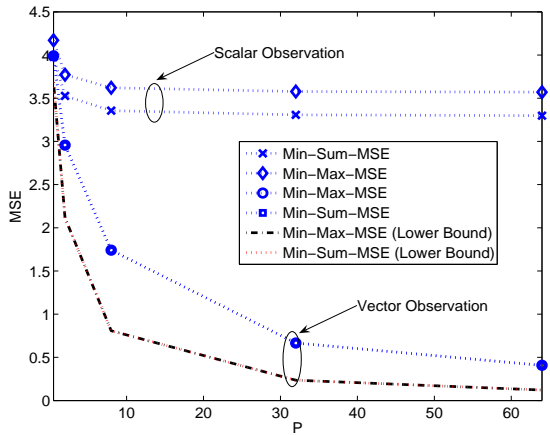}
\caption{Sum MSE vs. the value of constraint $P$}\label{f1}
\end{figure}

\begin{figure}[!htb]
\centering
\includegraphics[height=2.5in, width=3.3in]{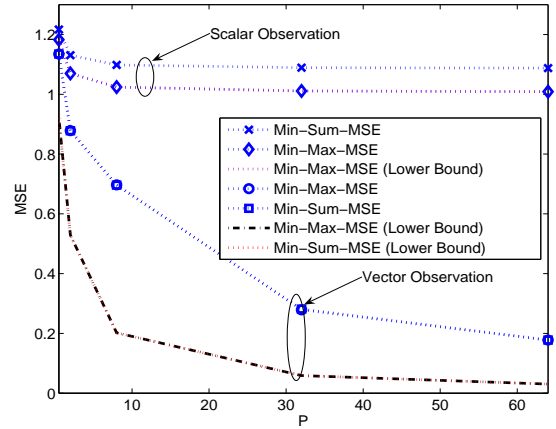}
\caption{Maximum MSE vs. the value of constraint $P$}\label{f2}
\end{figure}
\section{Simulations}\label{sec:four}
In the following simulations, the dimension of $\boldsymbol{\theta}_{n}$ and $\mathbf{a}$ are $M=4$ and $N=3$ respectively and $L$ is set to $4$.  The observation noise variance is set to $\sigma_v^2=0.5$,  and the covariance $\mathbf{Q}$ is assumed to be an identity matrix. The matrix $\mathbf{F}$ and the $\mathbf{G}$ are generated as complex Gaussian matrices with independent unit variance elements, and $\mathbf{F}$ is scaled to guarantee convergence of the Kalman filter. Once they are generated, $\mathbf{F}$ and $\mathbf{G}$ are kept constant in the simulation. The MSE performance for different constraints $P$ are calculated after convergence of the Kalman filter. In Figs.~\ref{f1}-\ref{f2}, for the vector-observation case the lower bound corresponds to the MSE calculated using $\mathbf{C}^{*}$. The performance gap between the lower bound and the proposed method represents the performance loss introduced by reducing the number of control parameters from $16$ to $3$. When the $P$ is small, the proposed method can achieve a performance close to the lower bound, \emph{e.g.}, when $P=0.5$ (corresponding to the first point on the curve), the performance degradation in the sum MSE is less than $10\%$. For the scalar observation case, we see as expected that the Min-Sum algorithm has the lowest sum-MSE and the Min-Max algorithm has the lowest maximum MSE.  Surprisingly, however, in the vector case the performance of the Min-Sum and Min-Max algorithms is essentially identical. The performance of the Min-Max-MSE algorithm appears to be equal to the lower bound which indicates that the solution $\mathbf{A}^{*}$ to problem (\ref{eq:minmaxsc6}) is very likely rank-one even with the constraint relaxed.  With increasing $P$, a performance floor exists for the scalar-observation case, while in the vector-observation case the MSE performance continues to improve.
\section{Conclusion}\label{sec:five}
In this paper, we investigated the problem of a Kalman filter with a linearly reconfigurable observation matrix. Two kinds of problems were formulated: Min-Sum-MSE or Min-Max-MSE. For the vector observation model, both of the optimization problems are difficult to solve directly, and we divided the problem into two simpler sub-problems that are easier to solve. Simulation results show that when the quadratic constraint is small, the proposed approach provides performance close to the MSE lower bound. For the scalar observation model, the Min-Sum-MSE problem is converted to a Rayleigh quotient maximization problem, for which an optimal closed-form solution is obtained, and for the Min-Max-MSE problem, we relax the rank-one constraint on the observation parameters and transform the optimization problem to an SDP feasibility problem. Based on the solution to the SDP feasibility problem, a rank-one solution can be reconstructed. Simulation results show that with a very high probability the solution of the relaxed problem is indeed rank-one. 
% Below is an example of how to insert images. Delete the ``\vspace'' line,
% uncomment the preceding line ``\centerline...'' and replace ``imageX.ps''
% with a suitable PostScript file name.
% -------------------------------------------------------------------------



\bibliographystyle{IEEEbib}

\bibliography{reference}
\end{document}